\documentclass[12pt]{article}
\usepackage{epsfig}
\newcommand{\ind}{$^{115}$In }
\newcommand{\tin}{$^{115}$Sn }
\newcommand{\isome}{$^{115m}$In }

\begin{document}

\begin{center}

{\Large {\bf Beta decay of $^{115}$In to the first excited level of $^{115}$Sn: \\
potential outcome for neutrino mass}}

\vskip 0.4cm

{\bf
C.M.~Cattadori$^{a,b}$,
M.~De~Deo$^a$,
M.~Laubenstein$^a$,
L.~Pandola$^a$,
V.I.~Tretyak$^{c,}$\footnote{Talk at 5th International Conference on
Non-Accelerator New Physics (NANP'05), Dubna, Russia, 20--25 June 2005.}
}

\vskip 0.4cm

{$^a$~\it INFN, Lab. Nazionali del Gran Sasso, S.S. 17 bis
km 18+910, I-67010 L'Aquila, Italy}

{$^b$~\it INFN Milano, Via Celoria 16, I-20133 Milano, Italy}

{$^c$~\it Institute for Nuclear Research, MSP 03680 Kiev, Ukraine}

\end{center}

\begin{abstract}
\noindent
Recent observation of $\beta$ decay of $^{115}$In to the first excited
level of $^{115}$Sn with an extremely low $Q_\beta$ value
($Q_\beta \le O(1)$ keV) could be used to set a limit on neutrino mass.
To give restriction potentially competitive with those
extracted from experiments with $^3$H ($\simeq$2 eV) and $^{187}$Re
($\simeq$15 eV), atomic mass difference between $^{115}$In and $^{115}$Sn
and energy of the first $^{115}$Sn level should be remeasured
with higher accuracy (possibly of the order of $\sim$1 eV).
\end{abstract}

\section{Introduction}

Development of new real-time solar neutrino detectors is of great interest
for current particle physics \cite{McD04}.
$^{115}$In was proposed long ago \cite{Rag76} as a promising target for solar neutrino
spectroscopy, having
low threshold of 114 keV for the $\nu_e$ capture \cite{ToI98}, which allows to
measure flux of low energy solar $pp$ neutrinos, and
high natural abundance of 95.71\% \cite{Ros98}.
The process of $\nu_e$ capture can be effectively discriminated from the background
processes using a specific tag: the emission of a prompt electron after the
$\nu_e$ capture $^{115}$In + $\nu_e$ $\to$ $^{115}$Sn($E_{exc2}$=613 keV) + $e^-$
with the subsequent
emission, after typical time delay of $\tau$=4.7~$\mu$s,
of two $\gamma$ quanta with energies of $E_{\gamma1}$=116 keV and
$E_{\gamma2}$=497 keV from deexcitation of the second excited level of $^{115}$Sn.
Notwithstanding these attractive features, the building of a
$^{115}$In-based detector
is a challenging task because $^{115}$In is unstable: it $\beta$ decays to ground state of
$^{115}$Sn (albeit with big half life of $T_{1/2}$=4.41$\times10^{14}$ yr \cite{Pfe79})
creating intensive irremovable background. This makes necessary to divide the detector
in small cells and search not only for time- but also for space-correlation
between the emitted electron and the gamma quanta.

The possibility to create a solar neutrino detector with $^{115}$In as a target is
under investigation in the LENS (Low Energy Neutrino Spectroscopy) project \cite{LENS}.
In this framework, in particular, In radiopurity and bremsstrahlung from beta decay
$^{115}$In $\to$ $^{115}$Sn were investigated, being important characteristics
which could prevent the successful exploitation of time- and space-correlations.
Measurements of In sample with HP Ge detectors were performed deep underground,
in the Gran Sasso National Laboratories (Italy), on the depth of 3800 m w.e.
As a by-product of these measurements, the $\beta$ decay of $^{115}$In to the first excited
level of $^{115}$Sn ($E_{exc1}$=497.4 keV) was observed at the first time \cite{Cat05}.
It has an extremely low intensity (1.2$\times10^{-6}$ in comparison with the $\beta$ decay
to $^{115}$Sn ground state) and long half life of $T_{1/2}$=3.7$\times10^{20}$ yr \cite{Cat05}.
These extreme values are related with the very low energy release, $Q_\beta$=1.6$\pm$4.0 keV,
that makes this process the $\beta$ decay with probably the lowest known $Q_\beta$ value.

After brief summary of the experiment and data analysis, we discuss in this paper a
possible use of the $^{115}$In $\to$ $^{115}$Sn($E_{exc1}$=497 keV) decay for setting a
limit on the neutrino mass.

\section{Detectors and measurements}

A sample of natural high purity In with weight of 928.7$\pm$0.1 g
was measured with four HP-Ge detectors mounted in one cryostat with a well in the centre.
The HP-Ge detectors were of 225.2, 225.0, 225.0 and 220.7 cm$^3$ volume, and had
typical energy resolution of 2.0 keV (FWHM) at the 1332 keV line
of $^{60}$Co. The experimental set-up was enclosed in a lead and copper passive shielding and
had a nitrogen ventilation system against radon. Data were collected in the Gran Sasso
National Laboratories (Italy), on the depth of 3800 m w.e. during 2762.3 h for the In sample
and during 1601.0 h for the background, in both cases with the complete shielding
around the detectors.

Statistics in the In and background measurements was accumulated in few independent
runs that resulted in some minor shifts in position of peaks present in the spectra.
Internal peaks of known origin and with good statistics were used to recalibrate
spectra from individual runs to obtain summary spectra with the help of the SAND0 routine
\cite{Tre90}. In result, the peaks positions in the sum spectra deviate from their
table values of less than 0.1 keV in the range of 300--2615 keV
both for the In and background measurements.

Efficiency of the detectors for gamma quanta emitted from the In sample was calculated
with a GEANT4-based code \cite{Cre00}. The results were checked in measurements with a
$^{60}$Co source, performed with the same set-up. The measured absolute efficiencies
agree with the computed ones within 12\% and are consistent within their statistical
uncertainties. In the following, we estimate the systematic uncertainty of the Monte
Carlo efficiencies to be 10\%.

The measured spectra for the In sample and for the background are presented in Fig.~1.
The bremsstrahlung emission from the $^{115}$In $\beta$ decay with end point of 499 keV
is clearly visible as the continuous component in the In spectrum.
In both spectra, 42 gamma lines with energy above 200 keV were
found. All lines (except of line with energy of 497.4 keV in the In sample)
were identified; they come from the natural radionuclides and
radioactive series ($^{40}$K, $^{238}$U, $^{235}$U, $^{232}$Th) and from cosmogenic
or antropogenic nuclides ($^{60}$Co, $^{137}$Cs, $^{207}$Bi, $^{26}$Al) that are
usually present as contaminations in copper and lead \cite{Cat05}. The counting rates of the
$\gamma$ lines for the In sample and the background were equal within their statistical
uncertainties.

\nopagebreak
\begin{figure}[ht]
\begin{center}
\mbox{\epsfig{figure=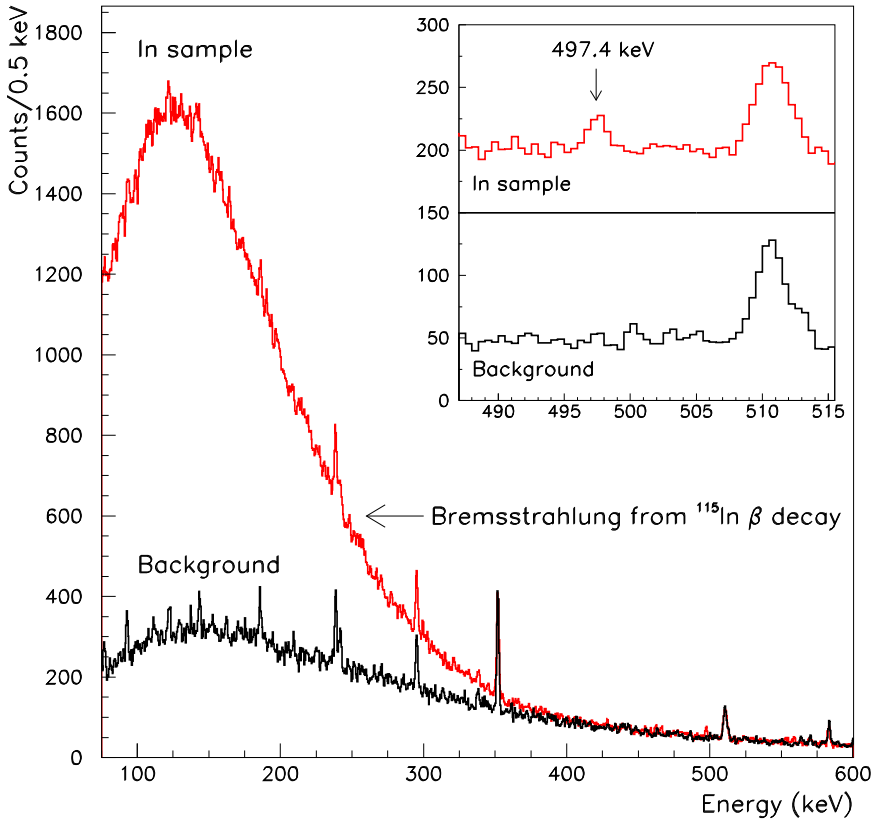,height=12.5cm}}
\caption{Experimental spectrum of the In sample (accumulated during 2762.3 h) and
background spectrum (1601.0 h) measured with 4 HP-Ge detectors at LNGS in
the energy interval 70--600 keV. The region of 600--2800 keV, where the spectra are practically
undistinguishable, is not shown. Background is normalized to the same counting time.
In the insert, the region of the 497.4 keV peak is shown in more detail; here the In spectrum
is shifted upward on 150 counts.}
\end{center}
\end{figure}

The only $\gamma$ line of the In spectrum which is not present in the background
measurement and cannot be ascribed to the usual radioactive contaminants is located
at the energy of 497.48$\pm$0.21 keV (see insert in Fig.~1).
>From the fit of the In spectrum in the energy region 487--508 keV with a Gaussian
peak and linear background assumption, the net area is 90$\pm$22 counts, inconsistent
with zero at more than 4$\sigma$. Variations of the energy interval for the fit result
in changes of the area inside the quoted uncertainty.
With the same procedure applied to the background spectrum, no Gaussian peak could be found,
and the resulting area is 0$\pm$14 counts; the corresponding upper limit derived with the
Feldman-Cousins method \cite{Fel98} is 23 counts at 90$\%$ C.L.
It can hence be concluded that the peak of 497.4 keV is statistically significant and
related with the In sample, being absent in the background measurement.

\section{Data analysis and interpretation}

The energy of the first excited level of $^{115}$Sn, daughter nucleus after
the $^{115}$In $\beta$ decay, is equal to 497.35$\pm$0.08 keV.
If populated, this level deexcites with the emission of a $\gamma$ quantum with energy
$E_\gamma$=497.358$\pm$0.024 keV \cite{ToI98,Bla99} which is in nice agreement
with that of the observed peak 497.48$\pm$0.21 keV.
However, the $Q_\beta$ value (i.e. atomic mass difference $\Delta M_a$ between
$^{115}$In and $^{115}$Sn) given in the atomic mass tables of Audi \& Wapstra,
known at the moment of our measurements, was equal to 495 keV \cite{Aud93} and
496 keV \cite{Aud95}. Thus the transition to the first excited level of $^{115}$Sn was
energetically forbidden\footnote{If to forget about 4 keV uncertainty in the $Q_\beta$ value
\cite{Aud93,Aud95}.}, and the $\beta$ decay of \ind was considered
as going exclusively to the ground state of \tin \cite{ToI98,Bla99} (see Fig.~2a).

\nopagebreak
\begin{figure}[ht]
\begin{center}
\mbox{\epsfig{figure=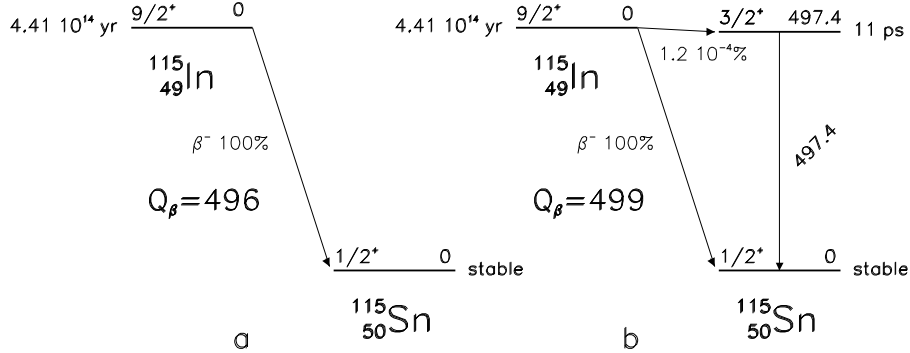,width=12.0cm}}
\caption{Old (a) and new (b) schemes of $^{115}$In $\to$ $^{115}$Sn
$\beta$ decay.}
\end{center}
\end{figure}

However, revised value of $Q_\beta$=499$\pm$4 keV in the last tables \cite{Aud03a} allowed
to avoid this contradiction: with this value, the decay to the first excited level
\ind $\to ^{115}$Sn($E_{exc1}$=497.4 keV) is kinematically allowed, though with an
extremely small $Q_\beta$ value, $Q_\beta$=1.6$\pm$4.0 keV.

Using the area of the 497 keV peak observed in the indium spectrum, corresponding partial
half life for the transition to the first excited level of \tin can be calculated as
\begin{equation}
T_{1/2}(^{115}\mbox{In} \to ^{115}\mbox{Sn}^*) =
\frac{\ln 2 \cdot N \cdot \varepsilon \cdot t}{S \cdot (1+\alpha)},
\end{equation}
where $N$ is the number of \ind nuclei in the sample, $\varepsilon$ is the efficiency to
detect the full energy $\gamma$ with the 4 HP Ge detectors, $t$ is the measurement time,
$S$ is the area of the peak and $\alpha$ is the coefficient of conversion of $\gamma$
quanta to electrons for the given nuclear transition.

The full peak efficiency at 497 keV was calculated with the Monte Carlo simulation
as $\varepsilon$=3.32$\pm$0.33\%.
Taking into account the total mass of the indium sample (928.7 g),
the atomic weight of indium (114.818 g$\cdot$mol$^{-1}$ \cite{Lae03}) and
the isotopic abundance of \ind (95.71\% \cite{Ros98}),
the number of \ind nuclei in our sample is $N$=4.66$\times 10^{24}$.
With the area of the peak 90$\pm$22 counts, the electron conversion coefficient
for the transition $\alpha$=8.1$\times 10^{-3}$ \cite{Bla99},
and $t$=2762.3 h, the $T_{1/2}$ value is equal
\begin{equation}
T_{1/2}(^{115}\textrm{In} \to ^{115}\textrm{Sn}^*) = (3.73\pm0.98)\times 10^{20}~\textrm{yr}.
\end{equation}

Half life for the ground state to ground state $\beta$ decay of $^{115}$In,
because of the large change in the nuclear angular momentum (9/2$^+$ $\to$ 1/2$^{+}$,
that classifies this decay as 4-fold forbidden transition) and of the
relatively small $Q_\beta$ value,
is equal to $T_{1/2}$=4.41$\times 10^{14}$ yr \cite{Pfe79,ToI98,Bla99,Aud03b}.
Thus the probability of decay to the first excited level is near one million times lower
than for the transition to the ground state of $^{115}$Sn; the experimental
branching ratio is $b$=(1.18$\pm$0.31)$\times 10^{-6}$.

The uncertainty on the half life and on the branching ratio mainly comes from the
statistical error on the net area of the 497 keV peak. An updated scheme of the
$^{115}$In $\to$ $^{115}$Sn $\beta$ decay is presented in Fig.~2b.

\section{Possible imitation of the effect}

In some nuclear processes $\gamma$ rays with energies close to 497 keV are emitted. This
could give an alternative explanation of the peak observed in the experimental spectrum.
Luckily, additional $\gamma$ rays are also emitted in such decays, allowing to tag those
mimicking effects.

The \ind nucleus has an isomeric state \isome with the energy $E_{iso}$=336.2 keV and
a half life of 4.5 h \cite{ToI98}. With the probability of 0.047\% the \isome
nucleus $\beta$ decays to the first excited level of $^{115}$Sn, with the subsequent
emission of a 497 keV $\gamma$ ray \cite{ToI98,Bla99}.
However, in this case a $\gamma$ ray with the energy $E_{iso}$=336.2 keV is emitted
with much higher probability (45.84\% \cite{ToI98}) because of the electromagnetic
transition from the isomeric \isome to the ground \ind state.
This huge peak at 336.2 keV, whose area should be $\sim$10$^3$
times bigger than that of the observed 497.4 keV peak, is absent in the
experimental spectrum; only a peak at 338.3 keV is observed, with the net area of
138$\pm$50 counts, which corresponds to the decay of $^{228}$Ac from the $^{232}$Th
natural chain. Therefore the decay of the isomeric state \isome is absolutely negligible
and the 497 keV peak cannot be ascribed to it, not even in part.

Capture of thermal neutrons by $^{115}$In results in $^{116}$In in excited state and
subsequent cascade of $\gamma$ quanta. In particular, $\gamma$ quantum with energy of
497.7 keV and relative yield of 1.32\%\footnote{Intensity of $\gamma$ line with energy of
273.0 keV is accepted as 100\%.} will be emitted \cite{capgam}. However,
accompanying intensive peak of 273.0 keV and, for example, peak of 492.5 keV (8.68\%)
are absent in the spectrum (see Fig.~1).
Capture of thermal $n$ on $^{113}$In (natural abundance 4.29\% \cite{Ros98})
also results in emission of $\gamma$ quantum with close energy: 496.7 keV (0.75\%) \cite{capgam}.
However, more intensive peaks (311.6 keV -- 100\%, 502.6 keV -- 5.00\%)
also are absent in the spectrum.
Thus, due to the underground location of the experimental setup and the low
flux of neutrons \cite{Bel89}, we conclude that $(n,\gamma)$ reactions cannot
contribute to the peak under analysis.

Protons produced by fast neutron or cosmic ray muons can populate the second excited level
of \tin ($E_{exc2}$=612.8 keV) via the $(p,n)$ reaction on \ind ($E_{thr}$=0.9 MeV);
the \tin nucleus quickly returns to the ground state with the emission of two $\gamma$
rays of energy 115.4 and 497.4 keV. The contribution originated by fast neutrons is
practically zero (see f.i. \cite{Cri95}) because of the deep
underground location and the lack of hydrogenous materials in the setup.
On the other hand, since the muon flux in the laboratory is extremely low
(1 $\mu$/(m$^2$$\cdot$h) \cite{Amb95}), also the contribution
of $(p,n)$ reactions induced by cosmic rays (see also \cite{Cri97}) to the 497 keV peak
is negligible ($<$10$^{-3}$ counts).

Some decays from the natural $^{238}$U and $^{232}$Th chains can also give $\gamma$ rays
in the energy region of interest, though with very low intensity.
They are in particular \cite{ToI98}
$^{214}$Bi ($E$ = 496.7 keV, $I$ = 0.0069\%),
$^{228}$Ac ($E$ = 497.5 keV, $I$ = 0.0059\%) and
$^{234m}$Pa ($E$ = 498.0 keV, $I$ = 0.062\%).
However, the sum contribution of these decays to the 497 keV peak is less than 1 count
and can be estimated using their stronger associated $\gamma$ lines. For instance,
the area of the 338.3 keV line of $^{228}$Ac, whose relative intensity is 11.27\%,
is only 138$\pm$50 counts. Therefore, if the contamination were located in the In sample,
the estimated contribution to the 497 keV peak, taking also into account the different
full peak efficiency, would be (7.3$\pm$2.6)$\times$10$^{-2}$ counts.

\section{$\beta$, $\gamma$-$\beta$ and $\beta$-$\gamma$ decays of $^{115}$In}

In addition, possible imitation of the effect could come from the so-called
$\gamma$-$\beta$ decay of $^{115}$In. In this process, also called induced
``photobeta" decay \cite{Sha65}, an external $\gamma$ quantum is absorbed by the nucleus,
thereby providing additional energy. This allows the $\beta$ transition to excited levels of
daughter nucleus, otherwise energetically forbidden, or even stimulates the $\beta$
decay of stable nuclei. In cases when the decay to ground state is strongly suppressed
by a large change in the spin (as for $^{115}$In $\to$ $^{115}$Sn), transitions
to excited levels would be preferable.

The process could be actual in the case of strong electromagnetic fields in stars
\cite{Sha65} or in a field of synchrotron radiation \cite{Kop05}, but also in the
searches for extremely rare processes, like ours. If the $^{115}$In nucleus absorbs a
$\gamma$ quantum with energy $E_\gamma$$>$114 keV from an external source, or even a bremsstrahlung
$\gamma$ from $^{115}$In $\beta$ decay itself, the second excited level of $^{115}$Sn, with
energy 612.8 keV, could be populated (spin changes from 9/2$^+$ to 7/2$^+$, and
this is an allowed $\beta$ transition). In the subsequent deexcitation process, two
$\gamma$'s with energies of 115.4 keV and 497.4 keV will be emitted, thus leading to the
peak of 497.4 keV (see Fig.~3).

\nopagebreak
\begin{figure}[ht]
\begin{center}
\mbox{\epsfig{figure=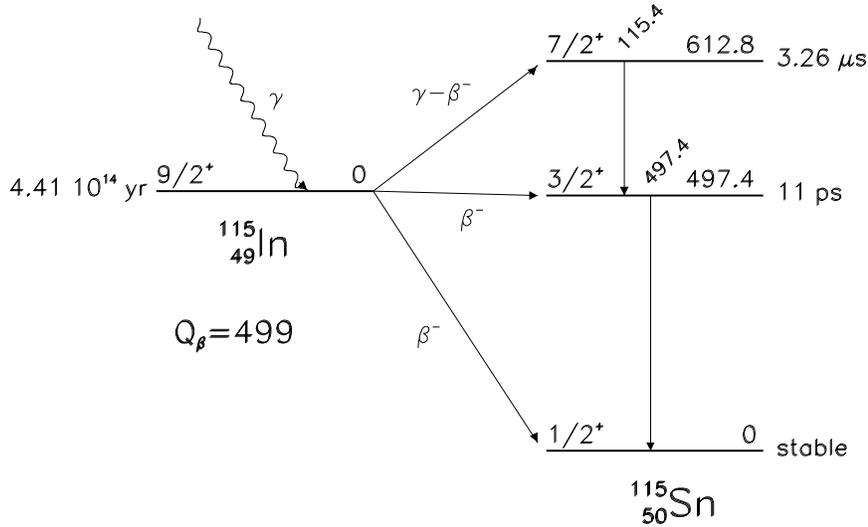,height=7.0cm}}
\caption{$\gamma$-$\beta$ decay of $^{115}$In $\to$ $^{115}$Sn. The absorption of the
external $\gamma$ quantum increases the nucleus energy and makes possible the $\beta$ decay
to the excited levels of $^{115}$Sn, otherwise energetically forbidden.}
\end{center}
\end{figure}

The calculated cross-sections of the $\gamma$-$\beta$ process are quite low: of the order of
10$^{-49}$--10$^{-45}$ cm$^{2}$, depending on the $Z$ of the parent nucleus, on $E_\gamma$ and
on the energy threshold \cite{Kop05}. Nevertheless, it was calculated that in a field of
intensive synchrotron radiation from the SPring-8 source (Japan), the $\beta$ decay of
$^{115}$In could go faster by 2 orders of magnitude \cite{Kop05}.
However, scaling the SPring-8 intensity of $\sim$10$^{17}$
$\gamma$/(s$\cdot$mm$^2$$\cdot$mrad$^2$$\cdot$keV) to the $\gamma$ radiation intensity
in our measurements, $<$0.1 $\gamma$/(s$\cdot$keV) (see Fig.~1), it is evident
that the contribution of $\gamma$-$\beta$ decay to the 497.4 keV peak is negligible.

It is interesting to note here another interesting process, the so-called
$\beta$-$\gamma$ decay \cite{Lon49}: when the direct $\beta$ transition
to the ground state of daughter nucleus is highly forbidden, a process
could take place in which a $\gamma$ quantum is emitted simultaneously with
the electron and antineutrino. Such a process is second-order in perturbation theory
and is similar to the double $\beta$ decay of a nucleus. The branching ratio of such
decay for $^{115}$In was calculated in \cite{Pac87} as: 6.7$\times 10^{-7}$.
Because in this case three particles ($\gamma$, $\beta$ and $\bar{\nu_e}$)
are emitted with
sum energy equal to $Q_\beta$, and the $\gamma$'s energy distribution is continuous,
the 497.4 keV peak could not be imitated. However, it should be noted that it could be
an additional source of $\gamma$ quanta in the In-based solar neutrino detector
(with the branching ratio of $\simeq$10$^{-6}$, as in $\beta$ decay $^{115}$In
$\to$ $^{115}$Sn$^*$) which also should be taken into account \cite{Pac87}.

\section{Possible outcome for the neutrino mass}

With the value of $Q_\beta$=1.6$\pm$4.0 keV, decay $^{115}$In $\to$ $^{115}$Sn$^*$
possibly is the $\beta$ decay with the lowest known $Q_\beta$ value (to be compared with that of
$^{163}$Ho: 2.555 keV and $^{187}$Re: 2.469 keV \cite{Aud03a}).
Below we will try to determine the $Q_\beta$ value more exactly on the basis of the
systematics of $\log ft$ values for such kind of decay and the measured value of $T_{1/2}$.

Nuclear spin and parity are changed in the observed transition
from the initial 9/2$^+$ of the \ind ground state to 3/2$^+$ of $^{115}$Sn$^*$;
this is therefore a 2-fold forbidden unique $\beta$ decay.
The recent compilation of $\log ft$ values \cite{Sin98} gives for
such a decay the average value $\log ft$=15.6$\pm$1.2;
for the 12 known experimental cases, the range is from 13.9 to 18.0.
With the measured value of the half life of (3.73$\pm$0.98)$\times 10^{20}$ yr,
the ``experimental'' $\log f$ value is $\log f$=--12.47$\pm$1.21.
On the other hand, the $\log f$ can be estimated with the help of the LOGFT tool
at the National Nuclear Data Center, USA \cite{logft} which is based on
the procedure described in \cite{Gov71}. For $Q_\beta$=1.6 keV, the value
calculated with the LOGFT code is $\log f$=--10.8;
this means that with such a $Q_\beta$ the $\beta$ decay should go near
50 times faster.

One can solve the inverse problem and use the LOGFT code to
adjust the $Q_\beta$ value corresponding to the ``experimental''
$\log f$=--12.47$\pm$1.21. Such a procedure gives a value of
$Q_\beta$=$460^{+700}_{-280}$ eV.
The lowest value of $\log ft$=13.9 in the range of the known
2-fold forbidden unique $\beta$ decays \cite{Sin98} corresponds to
$Q_\beta$=120 eV, while the highest value ($\log ft$=18.0) gives
$Q_\beta$=2.85 keV \footnote{We can derive on this basis
the atomic mass difference $^{115}$In--$^{115}$Sn: it
is equal to $497.9^{+2.3}_{-0.4}$ keV (with the error bars corresponding
to the whole range of 13.9--18.0 of $\log ft$ values),
which is more precise than the recent value of 499$\pm$4 keV \cite{Aud03a}.}.

While possibly the LOGFT tool was not intented to be applied for such low energies,
in any case it is clear that the $Q_\beta$ value in the $\beta$ decay
$^{115}$In $\to$ $^{115}$Sn$^*$ is very close to zero\footnote{Even the history
of the $Q_\beta$ evaluation for $^{115}$In gives some indication for this:
the $Q_\beta$ value was slightly lower than 497.4 keV energy of the first excited
$^{115}$Sn state in accordance with older tables of atomic masses,
$Q_\beta$=495$\pm$4 keV \cite{Aud93} and 496$\pm$4 keV \cite{Aud95},
while it is slightly higher in the last evaluation, 499$\pm$4 keV \cite{Aud03a}.}.
Such a unique situation could be used to establish a limit on the antineutrino mass,
in addition to the experiments with $^{3}$H and $^{187}$Re, where up-to-date limits
are in the range of $\simeq$2 eV \cite{Lob03} and $\simeq$15 eV \cite{Sis04},
respectively. Two approaches could be proposed.

(1) To measure the shape of the $\beta$ spectrum $^{115}$In $\to$ $^{115}$Sn$^*$,
registering $\beta$ particle in coincidence with $\gamma$ 497 keV, which allows
to reduce the background due to the $\beta$ decay of \ind to the ground state
of $^{115}$Sn. New In-based semiconductor detectors or fast bolometers could
be used for the purpose. Mass of antineutrino could be derived from distortion
of the spectrum shape (as in $^3$H experiments). It should be noted, however,
that such measurements would be quite difficult because of:
(i) extremely low $Q_\beta$ value;
(ii) the shape of 2-fold forbidden $\beta$ decay has to be theoretically
calculated very precisely, that may be not so easy also due to low $Q_\beta$ value.

(2) Just to use evident relation $m_\nu$$<$$Q_\beta$. Thus, low $Q_\beta$ value
means also low limit on neutrino mass. Because predictive power of any theoretical
calculation for such low energies is uncertain, the best way to derive potentially good
limit on $m_\nu$ is to measure experimentally with accuracy better than current:
(i) the atomic mass difference $\Delta M_a(^{115}$In--$^{115}$Sn);
(ii) the energy of the first excited level of $^{115}$Sn.

Energy of $\gamma$ quantum emitted from the first excited $^{115}$Sn level is
known currently as 497.358$\pm$0.024 keV \cite{ToI98,Bla99}, i.e. with precision of 24 eV.
However, this uncertainty could be reduced further in measurements of electron capture
$^{115}$Sb $\to$ $^{115}$Sn. It should be noted that many calibration lines of
radioactive sources were already measured with accuracy of 0.1--0.3 eV,
also in the $\sim$500 keV region of our interest \cite{ToI98}.

Current uncertainty $\delta(\Delta M_a)$ on atomic mass difference between
$^{115}$In and $^{115}$Sn is equal to 4 keV \cite{Aud03a}.
It should be noted that current technics (Penning traps) allows to reach accuracy
of $\le$10$^{-10}$ in atomic mass measurements \cite{DiF94,Bra99,Shi05}, that corresponds to
$\simeq$10 eV for nuclei with $A$$\simeq$100. For example, in work \cite{Shi05} absolute masses of
$^{32}$S and $^{33}$S were determined with accuracy of 1.5 eV, and masses of $^{129}$Xe and $^{132}$Xe
with accuracy of 9 eV. Even lower uncertainties could be expected for measurement of difference of
atomic masses.
As the first step, it could be useful to measure $\Delta M_a(^{115}$In--$^{115}$Sn)
with not-challenging accuracy of $\simeq$100 eV: if, for example, $\Delta M_a$
will be equal 460$\pm$100 eV, it will be enough not to expect good limit on $m_\nu$.
However, if it will be measured as $\simeq$0$\pm$100 eV, uncertainty on $\Delta M_a$
should be further reduced. In case if we are lucky and
$\Delta M_a(^{115}$In--$^{115}$Sn) $\simeq$ $E_{exc1}$($^{115}$Sn),
we could obtain $\lim m_\nu$ possibly concurrent with that from experiments with
$^3$H ($\simeq$2 eV) or $^{187}$Re ($\simeq$15 eV).
Both measurements of $E_{exc1}$ and $\Delta M_a(^{115}$In--$^{115}$Sn)
require strong experimental efforts but the physical result could be very
interesting and important.

\section{Conclusions}

Evidence for the previously unknown $\beta$ decay of \ind to the first excited state
of $^{115}$Sn at 497.4 keV was found from the measurement of the $\gamma$ spectrum
of a sample of metallic indium performed with HP Ge detectors in the
Gran Sasso Laboratory. The $Q_{\beta}$ value for this channel is
$Q_\beta$=1.6$\pm$4.0 keV which could be the lowest of all the known $\beta$ decays.
The branching ratio is found to be $b$=(1.2$\pm$0.3)$\times 10^{-6}$.

With measured value of $T_{1/2}$=(3.7$\pm$1.0)$\times 10^{20}$ yr and
calculation of $\log f$ value with the LOGFT code, the derived atomic mass
difference between $^{115}$In and $^{115}$Sn is equal $497.9^{+2.3}_{-0.4}$ keV,
which is more exact than the recent value 499$\pm$4 keV \cite{Aud03a}.

The low value of $Q_\beta$ could be used to set limit on the neutrino mass.
In the case when
$\Delta M_a(^{115}$In--$^{115}$Sn) $\simeq$ $E_{exc1}$($^{115}$Sn),
there is a chance to obtain result similar to that from
experiment with $^3$H ($\simeq$2 eV), if we will be able to
accurately measure $\Delta M_a(^{115}$In--$^{115}$Sn) and
$E_{exc1}$($^{115}$Sn), possibly with $\simeq$1 eV uncertainty.

~~~

One of us (V.I.T.) thanks organizers of the NANP'05 Conference for good organization and
nice conditions for work. He is also grateful to G. Audi, S. Rainville and E.G. Myers for
discussions on the atomic mass measurements.

\end{document}